\documentclass[journal]{IEEEtran}
\usepackage{subeqn}
\usepackage{graphicx}
\usepackage{float}
\usepackage{mathrsfs}
\usepackage{amsfonts}
\usepackage{cite}
\usepackage{color}
\usepackage{subfigure}
\usepackage{booktabs}
\usepackage{multirow}
\usepackage{url}

\usepackage[ruled,vlined]{algorithm2e}
\usepackage{algpseudocode}
\usepackage{graphics}

\ifCLASSINFOpdf
\else
\fi
\begin{document}

%%%%%%%%%%%%%%%%%%%%%%%%%%%%%%%%%%%%%%%%%%%%%%%%%%%%%%%%%%%%%%%%%%%%%%%%%%%%%
\title{Joint Phase Tracking and Channel Decoding for OFDM Physical-Layer Network Coding}
\author{Taotao Wang, \emph{Student Member, IEEE,} and Soung Chang Liew, \emph{Fellow, IEEE}
 % <-this % stops a space
%\thanks{This work is }    % <-this % stops a space
\thanks{The authors are with the Department of Information Engineering, The Chinese University of Hong Kong. Email: \{wtt011, soung\}@ie.cuhk.edu.hk   }% <-this % stops a space
}
% The paper headers
\markboth{Joint Phase Tracking and Channel Decoding for OFDM Physical-Layer Network Coding}%
{Shell \MakeLowercase{\textit{et al.}}: Bare Demo of IEEEtran.cls
for Journals}
% make the title area
\maketitle
%%%%%%%%%%%%%%%%%%%%%%%%%%%%%%%%%%%%%%%%%%%%%%%%%%%%%%%%%%%%%%%%%%%%%%%%%%%%%%
\vspace{-0.5in}

\begin{abstract}
%\boldmath
This paper investigates the problem of joint phase tracking and channel decoding in OFDM based Physical-layer Network Coding (PNC) systems. OFDM signaling can obviate the need for tight time synchronization among multiple simultaneous transmissions in the uplink of PNC systems. However, OFDM PNC systems are susceptible to phase drifts caused by residual carrier frequency offsets (CFOs). In the traditional OFDM system in which a receiver receives from only one transmitter, pilot tones are employed to aid phase tracking. In OFDM PNC systems, multiple transmitters transmit to a receiver, and these pilot tones must be shared among the multiple transmitters. This reduces the number of pilots that can be used by each transmitting node. Phase tracking in OFDM PNC is more challenging as a result. To overcome the degradation due to the reduced number of per-node pilots, this work supplements the pilots with the channel information contained in the data. In particular, we propose to solve the problems of phase tracking and channel decoding jointly. Our solution consists of the use of the expectation-maximization (EM) algorithm for phase tracking and the use of the belief propagation (BP) algorithm for channel decoding. The two problems are solved jointly through iterative processing between the EM and BP algorithms.  Simulations and real experiments based on software-defined radio show that the proposed method can improve phase tracking as well as  channel decoding performance.
\end{abstract}

\begin{IEEEkeywords}
Physical-layer network coding, expectation-maximization,  belief propagation, OFDM, phase tracking
\end{IEEEkeywords}

\IEEEpeerreviewmaketitle

\section{Introduction}

Relays can be employed to extend coverage, enhance reliability and increase throughput in wireless networks. Recently, the research community has shown growing interest in a simple relay network in which two terminal nodes communicate through a relay. This network is referred to as the two-way relay channel (TWRC). Physical-layer Network Coding (PNC), originally proposed in \cite{zhang2006hot}, can boost the throughput in TWRC by 100\% compared with the traditional multi-hop relaying method \cite{liew2011physical}. In TWRC operated with PNC, the two terminal nodes first transmit their messages simultaneously to the relay. The relay then maps the overlapped signals into a network-coded message (e.g., bit-wise XOR of the messages of the terminal nodes) and broadcasts the network-coded message to the two terminal nodes. Each terminal node then extracts the message of the other terminal node by subtracting its own message from the network-coded message. Thus, the two terminal nodes exchange one message with each other in two time slots. With traditional relaying, four time slots are needed \cite{liew2011physical}.

Although PNC has the potential to boost throughput in TWRC, there are new challenges for PNC. An important issue is how to perform PNC mapping at the relay when the two signals arrive with symbol and phase asynchronies. In \cite{lu2012asynchronous}, the authors devised a belief propagation (BP) \cite{kschischang2001factor, yedidia2003understanding} method to decode network-coded messages in asynchronous PNC systems. Single-carrier time-domain signals are assumed. Another solution for asynchronous PNC is to use OFDM signals \cite{rossetto2009design}. OFDM carries multiple data-streams on multiple subcarriers in a parallel manner, and thus the symbol duration within each subcarrier is lengthened compared to the single-carrier system. To deal with the effect of multipath, there is always a cyclic-prefix (CP) prepended at the beginning of each OFDM symbol. We designed and implemented an OFDM PNC system using software-defined radio (SDR) in \cite{lu2012implementation, lu2013real}. If the relative symbol delay between the two terminal nodes is within the CP, the time-domain misaligned samples will become aligned in the frequency-domain after DFT demodulation. This alleviates the strict synchronization requirement for PNC systems. Benefiting from the CP and the larger symbol duration, we can  perform PNC mapping one-by-one on each subcarrier in a manner similar to that in synchronous PNC.

One drawback of OFDM systems is their sensitivity to carrier frequency offset (CFO) between the crystal oscillators of the transmitter and receiver. CFO causes two negative effects: (i) the drifting of the phase of the channel coefficient of each subcarrier; (ii) the inter-carrier interferences (ICI) among different subcarriers. In systems such as 802.11 WLAN, we can estimate the CFO using preambles, and then compensate for the CFO for the whole packet. However, estimation error may leave behind an uncompensated residual CFO.  Since the residual CFO is relatively small, the CFO-induced ICI is typically small. We can treat the ICI as an additional noise which degrades the effective received signal-to-noise ratio (SNR), and correct it using powerful channel codes \cite{sathananathan2001forward}. However, even a small residual CFO can lead to large phase drifts which accumulate over time. Data decoding will fail if we ignore the phase drifts. It is therefore important to track phase drifts and compensate for them before data decoding.

For WLAN, it is common to track phase using pilot tones in the OFDM symbols. In this paper, we consider the more challenging phase tracking problem in OFDM PNC systems. The pilot tones are now shared by the two terminal nodes. This reduces the number of pilots that can be used by each node. Moreover, the superimposed constellation at the relay receiver is denser than that of a single user receiver.

In this work, we propose to tackle the problem of phase tracking jointly with channel coding in an iterative manner. In particular, we use the framework of expectation-maximization and belief propagation (EM-BP) \cite{embp2013techrep} to perform joint phase tracking and channel decoding. The BP algorithm performs channel decoding using the phase-adjusted CSI estimated by the EM algorithm, while the EM algorithm performs phase tracking using the pilot symbols as well as the soft information of data symbols provided by the BP algorithm.

Since the proposed EM-BP algorithm makes use of both the data symbols and pilot symbols for phase tracking, we can potentially improve the accuracy of phase tracking compared with the traditional pilot-based algorithm. We conduct simulations and experiments to evaluate the proposed EM-BP algorithm. Our experiments are based on the data collected from an OFDM PNC protoype \cite{lu2012implementation, lu2013real}. The experimental results show that the EM-BP algorithm can yield 2-3 dB gain in BER performance compared with the traditional method that performs pilot-based phase tracking and channel decoding in a disjoint manner.

\section{System Model}

\begin{figure}[!t]
\centering
\includegraphics[width=3.5in]{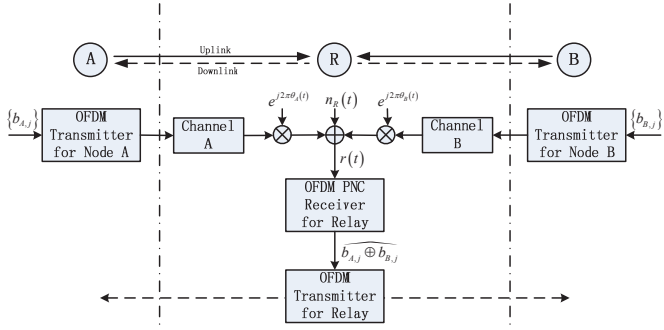}
\caption{The system model of OFDM PNC system.}
\end{figure}

\subsection{Transmit Signal}
Fig. 1 shows our system model. In the uplink of OFDM PNC, terminal nodes A and B simultaneously transmit signals to the relay node. We assume that the transmitters of nodes A and B employ the frame format proposed in \cite{lu2012implementation}, in which the preambles of A and B do not overlap, but the payloads overlap in time. One frame contains $M$  OFDM symbols for the payload. Let $N$ be the number of subcarriers. The length-$N$ vector ${{\bf{X}}_{u,m}} = {\left[ {\begin{array}{*{20}{c}}
   {{X_{u,m,0}}} & {{X_{u,m,1}}} &  \cdots  & {{X_{u,m,N - 1}}}  \\
\end{array}} \right]^T}$ is the $m^{th}$ frequency-domain symbol transmitted by node $u \in \left\{ {A,B} \right\}$. Within the $N$ subcarriers, each OFDM symbol has $N_d$ data tones, $N_p$   pilot tones and $N_z$  zero tones ($N = {N_d} + {N_p} + {N_z}$). The symbols on data tones are obtained after channel encoding, interleaving and constellation mapping. In this work, we assume that nodes A and B use the same channel encoder (the valid set of codewords is denoted by $C$), interleaver and constellation when mapping their source information bits $\left\{ {{b_{A,j}}} \right\}$, $\left\{ {{b_{B,j}}} \right\}$ to the transmitted data symbols. The pilots assist the task of phase tracking. We adopt a pilot pattern similar to that employed in \cite{lu2012implementation}, where separate pilot tones in the frequency domain are used by the two end nodes. The zero tones serve as the guard bands. The OFDM modulation is implemented with an $N$ point IDFT: ${{\bf{x}}_{u,m}} = IDFT\left( {{{\bf{X}}_{u,m}}} \right)$, where ${{\bf{x}}_{u,m}} = {\left[ {\begin{array}{*{20}{c}}
   {{x_{u,m,0}}} & {{x_{u,m,1}}} &  \cdots  & {{x_{u,m,N - 1}}}  \\
\end{array}} \right]^T}$ is the vector of time-domain samples. To combat the delay spread of multipath channels, which cause inter-symbol interference (ISI), the $N$ time-domain samples of one OFDM symbol is preceded by a CP. The length of the CP is $N_{cp}$. Therefore, each OFDM symbol corresponds to ${N_s} = {N_{cp}} + N$ time-domain samples, where ${x_{u,m,i}} = {x_{u,m,i + N}}$ for $i \in \left[ { - {N_{cp}}, - 1} \right]$. After pulse shaping and digital-to-analogy conversion (DAC), the ${N_s}$ time-domain samples of the $m^{th}$ OFDM symbol are converted into analog waveforms:
\begin{equation}
{s_{u,m}}\left( t \right) = \sum\limits_{i =  - {N_{cp}}}^{N - 1} {{x_{u,m,i}}\varphi \left( {t - i{T_s}} \right)}
\end{equation}
where $\varphi \left( t \right)$ is the shaping pulse for ensuring the transmitted signal satisfy the spectrum requirement, and $T_s$ is the sampling interval (The OFDM symbol duration is therefore$T=N_sT_s$).

The relay broadcasts a beacon to coordinate the uplink transmissions \cite{lu2013real}. The beacon accomplishes two functions: (i) First, it triggers simultaneous transmissions by the two terminal nodes to meet a loose time synchronization requirement (we will elaborate on this later). (ii) Second, it allows the two terminal nodes to do CFO estimation and compensate for the CFO by precoding in the uplink transmissions to the relay. Specifically, each node estimates the CFO between itself and the relay based on the received beacon signal; and then multiplies its signal with a compensating phase term before transmission \cite{lu2013real}. In this way, we can reduce the CFO to a small residual CFO.

\subsection{Receive Signal}
The received signal waveform at the relay can be expressed as
\begin{equation}
r\left( t \right) = \sum\limits_{u \in \left\{ {A,B} \right\}} {\sum\limits_{m = 0}^{M - 1} {{s_{u,m}}\left( {t - mT} \right) \otimes {h_u}\left( t \right){e^{j{\theta _u}\left( t \right)}}} }  + {n_R}\left( t \right)
\end{equation}
where $\otimes$ is the convolution operator, ${h_u}\left( t \right)$ is the frequency-selective channel between node $u$ and the relay,  ${{\theta _u}\left( t \right)}$ is the phase drift between the relay and node $u$  caused by the residual CFO,  ${n_R}\left( t \right)$ is the thermal noise from the receiver frontend modeled as a zero-mean Gaussian random process with variance $\sigma _n^2$.  We define the maximum delay spread of a multipath channel as the difference of the delay of the longest path and the delay of the shortest path, and denote the maximum delay spread of the frequency-selective channel of node $u$, ${h_u}\left( t \right)$, by ${\tau _u}$. If we use the arrival time of node A as the reference time and assume node B arrives later than node A (the difference between the delay of the shortest path of node B and the delay of the shortest path of node A is denoted by ${\tau}$), it can be shown \cite{lu2012implementation} that as long as the delay-spread-within-CP requirement, $\max \left\{ {{\tau _A},\tau  + {\tau _B}} \right\} \le {N_{cp}}{T_s}$, is satisfied, the time-domain symbol misalignment does not cause any negative effect on the carrier-by-carrier PNC mapping in the frequency-domain. This delay-spread-within-CP requirement can be regarded as a loose time synchronization requirement and can be achieved using beacon triggering \cite{lu2012implementation, lu2013real} of simultaneous transmissions by nodes A and B. Therefore, with OFDM signaling, the time asynchrony in PNC systems is not a critical issue. However, the above conclusion is made under the assumption that there is no phase drift in the system. The presence of phase drift will degrade the system performance.

After ADC and removal of CP, the $N$ discrete samples of the $m^{th}$  received OFDM symbol are collected into a vector  ${{\bf{r}}_m} = {\left[ {\begin{array}{*{20}{c}}
   {{r_{m,0}}} & {{r_{m,1}}} &  \cdots  & {{r_{m,N - 1}}}  \\
\end{array}} \right]^T}$. The frequency-domain sample vector of the $m^{th}$ OFDM symbol, ${{\bf{R}}_m} = {\left[ {\begin{array}{*{20}{c}}
   {{R_{m,0}}} & {{R_{m,1}}} &  \cdots  & {{R_{m,N - 1}}}  \\
\end{array}} \right]^T}$, is obtained by performing DFT on ${{\bf{r}}_m}$. We can express ${{\bf{R}}_m}$ as
\begin{equation}
\begin{array}{l}
 {{\bf{R}}_m}{\rm{ }} = \sum\limits_{u \in \left\{ {A,B} \right\}} {{e^{j{\Theta _{u,m}}}}{\bf{D}}\left( {{{\bf{X}}_{u,m}}} \right){{\bf{H}}_u}}  + \underbrace {\sum\limits_{u \in \left\{ {A,B} \right\}} {{{\bf{\Lambda }}_{u,m}}}  + {{\bf{N}}_{R,m}}}_{ \buildrel \Delta \over = {{\bf{W}}_{R,m}}} \\
 {\rm{      }} = \sum\limits_{u \in \left\{ {A,B} \right\}} {{e^{j{\Theta _{u,m}}}}{\bf{D}}\left( {{{\bf{X}}_{u,m}}} \right){{\bf{H}}_u}}  + {{\bf{W}}_{R,m}} \\
 \end{array}
\end{equation}
where  ${\Theta _{u,m}}$ is the phase drift on the $m^{th}$  frequency-domain OFDM symbol from node $u$;  ${\bf{D}}\left( {{{\bf{X}}_{u,m}}} \right) = diag\left( {{{\bf{X}}_{u,m}}} \right)$
is the diagonal matrix with transmitted symbols ${{\bf{X}}_{u,m}}$ as its diagonal elements; ${{\bf{H}}_u} = {\left[ {\begin{array}{*{20}{c}}
   {{H_{u,0}}} & {{H_{u,1}}} &  \cdots  & {{H_{u,N - 1}}}  \\
\end{array}} \right]^T}$ is the frequency response of the channel between node $u$  and the relay, ${{\bf{\Lambda }}_{u,m}}$ is the ICI component from node $u$, and ${{\bf{N}}_{R,m}}$
is the frequency-domain noise. In (3), ${{\bf{W}}_{R,m}}$ is the ICI plus noise.  In this paper, we approximate ${{\bf{W}}_{R,m}}$  as a circularly symmetric white Gaussian noise with zero mean and covariance matrix $\sigma _W^2{{\bf{I}}_N}$. Based on (3), we can apply the EM-BP algorithm to do joint phase tracking and channel decoding.

\section{EM-BP for OFDM PNC}
In this section, we derive the EM-BP algorithm for joint phase tracking and channel decoding in OFDM PNC systems. Define the pair of phase drifts on the $m^{th}$  OFDM symbol as ${{\bf{\Theta }}_m} \buildrel \Delta \over = {\left[ {\begin{array}{*{20}{c}}
   {{\Theta _{A,m}}} & {{\Theta _{B,m}}}  \\
\end{array}} \right]^T}$; the vector containing the phase drifts over all OFDM symbols as ${\bf{E}} = {\left[ {\begin{array}{*{20}{c}}
   {{\bf{\Theta }}_0^T} & {{\bf{\Theta }}_1^T} &  \cdots  & {{\bf{\Theta }}_{M - 1}^T}  \\
\end{array}} \right]^T}$; the pair of data symbols on the $i^{th}$ subcarrier of the $m^{th}$   OFDM symbol as ${{\bf{X}}_{m,i}} \buildrel \Delta \over = {\left[ {{X_{A,m,i}},{X_{B,m,i}}} \right]^T}$; and the vector containing all the received frequency-domain OFDM samples as ${\bf{R}} \buildrel \Delta \over = {\left[ {\begin{array}{*{20}{c}}
   {{\bf{R}}_0^T} & {{\bf{R}}_1^T} &  \cdots  & {{\bf{R}}_{M - 1}^T}  \\
\end{array}} \right]^T}$. We assume that the channels ${\left\{ {{{\bf{H}}_u}} \right\}_{u = \left\{ {A,B} \right\}}}$ are already known at the relay, which can be achieved via estimation using the preamble symbols.

EM-BP is an iterative framework where the $k^{th}$ iteration consists of a BP algorithm for channel decoding and an EM algorithm for phase tracking.

\subsection{BP for virtual channel decoding}
With the estimate of phase drift ${\widehat{\bf{E}}^{\left( {k - 1} \right)}} = {\left[ {\begin{array}{*{20}{c}}
   {\widehat{\bf{\Theta }}_0^{\left( {k - 1} \right)T}} & {\widehat{\bf{\Theta }}_1^{\left( {k - 1} \right)T}} &  \cdots  & {\widehat{\bf{\Theta }}_{M - 1}^{\left( {k - 1} \right)T}}  \\
\end{array}} \right]^T}$ from the $k-1^{th}$  iteration, we perform BP for channel decoding to find $p\left( {{{\bf{X}}_{m,i}}\left| {{\bf{R}},{{\widehat{\bf{E}}}^{\left( {k - 1} \right)}},{C^2}} \right.} \right)$ for all $m$  and all $i$, where $C^2$  is the code constraint imposed by a `virtual channel encoder' that takes the original information source symbols from nodes A and B $\left\{ {{b_{A,j}},{b_{B,j}}} \right\}$ as inputs, and outputs $\left\{ {{{\bf{X}}_{m,i}}} \right\}$ as coded symbols. The BP algorithm for virtual channel decoding applies the sum-product \cite{kschischang2001factor} rule on the factor graph that incorporates the constraints imposed by virtual channel encoding, which models the simultaneous transmissions by the two terminal nodes \cite{zhang2009channel, wubben2010generalized, lu2012asynchronous}. The BP virtual channel decoding is initialized with the probabilities $p\left( {{R_{m,i}}\left| {{{\bf{X}}_{m,i}},\widehat{\bf{\Theta }}_m^{\left( {k - 1} \right)}} \right.} \right)$ for all  $m$  and all $i$ as inputs. Here, $p\left( {{R_{m,i}}\left| {{{\bf{X}}_{m,i}},\widehat{\bf{\Theta }}_m^{\left( {k - 1} \right)}} \right.} \right)$ has a Gaussian form:
\begin{equation}
\begin{array}{l}
 p\left( {{R_{m,i}}\left| {{{\bf{X}}_{m,i}},\widehat{\bf{\Theta }}_m^{\left( {k - 1} \right)}} \right.} \right) \\
  \propto \exp \left( { - {{{{\left\| {{R_{m,i}} - \sum\limits_{u \in \left\{ {A,B} \right\}} {{e^{j\widehat\Theta _m^{\left( {k - 1} \right)}}}{X_{u,m,i}}{H_{u,i}}} } \right\|}^2}} \mathord{\left/
 {\vphantom {{{{\left\| {{R_{m,i}} - \sum\limits_{u \in \left\{ {A,B} \right\}} {{e^{j\widehat\Theta _m^{\left( {k - 1} \right)}}}{X_{u,m,i}}{H_{u,i}}} } \right\|}^2}} {\sigma _W^2}}} \right.
 \kern-\nulldelimiterspace} {\sigma _W^2}}} \right) \\
 \end{array}
\end{equation}
The decoding results of the BP algorithm is then fed to the EM algorithm for phase tracking.

\subsection{EM for Phase Tracking}
We update the ${k^{th}}$ estimate for the phase drifts ${\widehat{\bf{{\rm E}}}^{\left( k \right)}}$ according to the EM algorithm consisting of an E-step and an M-step \cite{dempster1977EM}.

\noindent \textbf{E-Step:}

We compute the $Q$ function of the phase drifts ${\bf{E}}$ given the last estimate ${\widehat{\bf{E}}^{\left( {k - 1} \right)}}$:
\begin{equation}
\begin{array}{l}
 Q\left( {{\bf{E}}\left| {{{\widehat{\bf{E}}}^{\left( {k - 1} \right)}}} \right.} \right) \\= \sum\limits_{{{\bf{X}}_A},{{\bf{X}}_B}} {\log p\left( {{\bf{R}}\left| {{{\bf{X}}_A},{{\bf{X}}_B},{\bf{E}}} \right.} \right)} p\left( {{{\bf{X}}_A},{{\bf{X}}_B}\left| {{\bf{R}},{{\widehat{\bf{E}}}^{\left( {k - 1} \right)}},{C^2}} \right.} \right) \\
 {\rm{                     = }}\sum\limits_m {\sum\limits_i {\sum\limits_{{{\bf{X}}_{m,i}}} {\log p\left( {{R_{m,i}}\left| {{{\bf{X}}_{m,i}},{\bf{E}}} \right.} \right)} } } p\left( {{{\bf{X}}_{m,i}}\left| {{\bf{R}},{{\widehat{\bf{E}}}^{\left( {k - 1} \right)}},{C^2}} \right.} \right) \\
 \end{array}
 \end{equation}
where ${{\bf{X}}_u} \buildrel \Delta \over = {\left[ {\begin{array}{*{20}{c}}
   {{\bf{X}}_{u,0}^T} & {{\bf{X}}_{u,1}^T} &  \cdots  & {{\bf{X}}_{u,M - 1}^T}  \\
\end{array}} \right]^T}$ is the vector containing all the symbols transmitted by node $u$.

Note that in (5) the outputs from the BP channel decoding $\left\{ {p\left( {{{\bf{X}}_{m,i}}\left| {{\bf{R}},{{\widehat{\bf{E}}}^{\left( {k - 1} \right)}},{C^2}} \right.} \right)} \right\}$ are used to assist EM phase tracking. To allow `decoupled' computation of (5), we make a simplification to ignore the relationship of the phase drift over symbols: that is, we assume  ${{\bf{\Theta }}_m}$ and ${{\bf{\Theta }}_{m'}}$ are independent for $m \ne m'$.  With this assumption, we can rewrite (5) as
\begin{equation}
Q\left( {{\bf{E}}\left| {{{\widehat{\bf{E}}}^{\left( {k - 1} \right)}}} \right.} \right) = \sum\limits_m {{Q_m}\left( {{{\bf{\Theta }}_m}\left| {\widehat{\bf{\Theta }}_m^{\left( {k - 1} \right)}} \right.} \right)}
\end{equation}
where ${Q_m}\left(  \cdot  \right)$ is the symbol-wise Q function of the   OFDM symbol:
\begin{equation}
\begin{array}{l}
 {Q_m}\left( {{{\bf{\Theta }}_m}\left| {\widehat{\bf{\Theta }}_m^{\left( {k - 1} \right)}} \right.} \right) \\= \sum\limits_i {\sum\limits_{{{\bf{X}}_{m,i}}} {\log p\left( {{R_{m,i}}\left| {{{\bf{X}}_{m,i}},{{\bf{\Theta }}_m}} \right.} \right)} } p\left( {{{\bf{X}}_{m,i}}\left| {{\bf{R}},{{\widehat{\bf{{\bf{E}}}}}^{\left( {k - 1} \right)}},{C^2}} \right.} \right) \\
 \propto  - \sum\limits_i {\sum\limits_{{{\bf{X}}_{m,i}}} {{{\left\| {{R_{m,i}} - \sum\limits_{u \in \left\{ {A,B} \right\}} {{e^{j{\Theta _{u,m}}}}{X_{u,m,i}}{H_{u,i}}} } \right\|}^2}} } \times \\
 \;\;\;\;\;\;\;\;\;\;\;\;\;\;\;\;\;\;\;\;\;\;\;\;\;\;\;\;\;\;\;\;\;\;\;\;\;\;\;\;\;\;\;\;\;\;\;\;\;\;\;\;  p\left( {{{\bf{X}}_{m,i}}\left| {{\bf{R}},{{\widehat{\bf{{\bf{E}}}}}^{\left( {k - 1} \right)}},{C^2}} \right.} \right) \\
 \end{array}
 \end{equation}
This decoupled expression for Q function can simplify the computation in the M-step of EM.

\noindent \textbf{M-Step:}

The objective of M-step is to find the variable that maximizes the Q function defined as in (5). Due to the simplification by (6), it is equivalent to finding
\begin{equation}
\widehat{\bf{\Theta }}_m^{\left( k \right)} = \arg \mathop {\max }\limits_{{{\bf{\Theta }}_m}} {Q_m}\left( {{{\bf{\Theta }}_m}\left| {\widehat{\bf{\Theta }}_m^{\left( {k - 1} \right)}} \right.} \right)
\end{equation}
for each $m$. This is a symbol-by-symbol phase tracking process. To solve (8), we use a particle-filtering \cite{arulampalam2002tutorial} type method to locate the argument that maximizes the Q function calculated in (7). We use a list of samples (called particles) and the associated weights (the function values of these particles) to represent the Q function. According to the weights, we re-compute a new set of particles that are adaptively closer to the peak location, and then iterate. This method is proposed in \cite{dauwels2004phase} to enable a practical message passing algorithm for the phase estimation in a single-user single-carrier system. Here we modify it only slightly for our purpose. The pseudo-code for particle-filtering type method that solves (8) can be found in Algorithm 1, where $P$ is the iteration number of particle-filtering, $L^2$ is the number of particles and  $\varepsilon$ is the forgetting factor. In our simulations and experiments in Section IV, we set $P=4$, $L=10$ and $\varepsilon=0.1$.

\begin{algorithm}
\caption{Particle-filtering method for solving the M-step of EM}
\KwIn{${\bf{R}}_m$, $\left\{ {{{\bf{H}}_u}} \right\}$, $\left\{ {p\left( {{{\bf{X}}_{m,i}}\left| {\bf{R}} \right.,{{\widehat{\bf{E}}}^{\left( {k - 1} \right)}},{C^2}} \right)} \right\}$}
\KwOut{$\widehat{\bf{\Theta }}_m^{\left( k \right)}$}
 initialize the list of $L^2$ samples as a $L \times L$ matrix ${{{\bf{\Theta }}^{\left( 0 \right)}}}$ whose ${\left( {p,q} \right)^{th}}$ element is given by ${\bf{\Theta }}_{p,q}^{\left( 0 \right)}={\left[ {\begin{array}{*{20}{c}}
   {p{{2\pi } \mathord{\left/
 {\vphantom {{2\pi } L}} \right.
 \kern-\nulldelimiterspace} L}} & {q{{2\pi } \mathord{\left/
 {\vphantom {{2\pi } L}} \right.
 \kern-\nulldelimiterspace} L}}  \\
\end{array}} \right]^T}$ for $0 \le p,q \le L - 1$\;

\For {$l =1$ to $P$}
{
compute the weights ${\widetilde\omega _{p,q}}= {Q_m}\left( {{\bf{\Theta }}_{p,q}^{\left( {l - 1} \right)}\left| {\widehat{\bf{\Theta }}_m^{\left( {k - 1} \right)}} \right.} \right)$, ${\omega _{p,q}} = \gamma {\widetilde\omega _{p,q}}$ for $0 \le p,q \le L - 1$\, where ${\gamma ^{ - 1}} = \sum\nolimits_p {\sum\nolimits_q {{{\widetilde\omega }_{p,q}}} }$\;

update the list of samples according to ${\bf{\Theta }}_{p,q}^{\left( l \right)} = \left( {1 - \varepsilon } \right){\bf{\Theta }}_{p,q}^{\left( {l - 1} \right)} + \varepsilon {\overline {\bf{\Theta }} ^{\left( {l - 1} \right)}}$ for $0 \le p,q \le L - 1$, where $\varepsilon$ is a forgetting factor and ${\overline {\bf{\Theta }} ^{\left( {l - 1} \right)}} = \sum\nolimits_p {\sum\nolimits_q {{\omega _{p,q}}{\bf{\Theta }}_{p,q}^{\left( {l - 1} \right)}} }$\;

}
$\widehat{\bf{\Theta }}_m^{\left( k \right)} = \arg \mathop {\max }\limits_{{\bf{\Theta }}_{p,q}^{\left( P \right)}} {Q_m}\left( {{\bf{\Theta }}_{p,q}^{\left( P \right)}\left| {\widehat{\bf{\Theta }}_m^{\left( {k - 1} \right)}} \right.} \right)$\;
return $\widehat{\bf{\Theta }}_m^{\left( k \right)}$\;
\end{algorithm}

We carry out the above EM phase tracking symbol by symbol. After updating the phase drifts of all the OFDM symbols, we obtain ${\widehat{\bf{E}}^{\left( k \right)}} = {\left[ {\begin{array}{*{20}{c}}
   {\widehat{\bf{\Theta }}_0^{\left( k \right)T}} & {\widehat{\bf{\Theta }}_1^{\left( k \right)T}} &  \cdots  & {\widehat{\bf{\Theta }}_{M - 1}^{\left( k \right)T}}  \\
\end{array}} \right]^T}$ and then iterate to perform the next BP channel decoding iteration.

\subsection{Initialization and Termination of BP-EM iteration}

The EM mechanism can usually find the maximum likelihood (ML) estimate of the phase upon convergence of the iterations. However, convergence is predicated on a good initial point  for the EM iterations \cite{wu1983convergence}. In this work, we choose to use the least square (LS) estimation \cite{kay1998fundamentals} of the phases from pilot tones in the $m^{th}$  OFDM symbol as the initial point $\widehat{\bf{\Theta }}_m^{\left( 0 \right)}$ for all $m$. We denote the set that contains the indexes of pilot tones assigned to node $u$ by $P_u$. Node $u$ just transmits known symbols on the pilot tones indexed by $P_u$ and null its signals on the other pilot tones. The initial phase estimates by LS pilot-based estimation are
\begin{equation}
\hat \Theta _{u,m}^{\left( 0 \right)} = \angle \left( {\sum\limits_{i \in {P_u}} {X_{u,m,i}^ * } {R_{m,i}}} \right)
\end{equation}
for $u \in \left\{ {A,B} \right\}$ and $m = 0,1, \cdots ,M - 1$, where $\angle \left(  \cdot  \right)$ is the angle of a complex signal, ${\left(  \cdot  \right)^*}$ is the conjugate operator.

We repeat BP channel decoding and EM phase tracking iteratively. When the number of iterations  reaches a preset maximum limit $K$, we terminate the BP-EM algorithm after obtaining the final phase estimate ${\widehat{\bf{{\bf{E}}}}^{\left( K \right)}} = {\left[ {\begin{array}{*{20}{c}}
   {\widehat{\bf{\Theta }}_0^{\left( K \right)T}} & {\widehat{\bf{\Theta }}_1^{\left( K \right)T}} &  \cdots  & {\widehat{\bf{\Theta }}_{M - 1}^{\left( K \right)T}}  \\
\end{array}} \right]^T}$.  Substituting $\widehat{\bf{\Theta }}_m^{\left( K \right)}$ into (4) to replace $\widehat{\bf{\Theta }}_m^{\left( {k - 1} \right)}$ for all $m$ and all $i$, we carry out a final round of BP channel decoding to obtain $p\left( {{b_{A,j}},{b_{B,j}}\left| {{\bf{R}},{{\widehat{\bf{{\bf{E}}}}}^{\left( K \right)}},{C^2}} \right.} \right)$ for all $j$. Then, the network-coded source message is obtained by
\begin{equation}
\begin{array}{l}
 \widehat{{b_{A,j}} \oplus {b_{B,j}}} \\
  = \arg \mathop {\max }\limits_b {\sum _{\begin{array}{*{20}{c}}
   {{b_{A,j}},{b_{B,j}}:}  \\
   {{b_{A,j}} \oplus {b_{B,j}} = b}  \\
\end{array}}}p\left( {{b_{A,j}},{b_{B,j}}\left| {{\bf{R}},{{\widehat{\bf{{\rm E}}}}^{\left( K \right)}},{C^2}} \right.} \right) \\
 \end{array}
\end{equation}
After that, the relay channel-encodes the network-coded source message and broadcasts the channel-coded message to nodes A and B in the downlink phase, as illustrated in Fig. 1.

\section{Simulations and Experimental Results}

This section presents computer simulations and experimental results for the evaluation of the proposed algorithm. The frame format used is the one proposed in \cite{lu2012implementation}, a slightly modified version of 802.11 frame format. The DFT size is $N = 64$. The CP length is $N_{cp} = 16$. One OFDM symbol includes $N_{d} = 48$ data tones, $N_{p} = 4$  pilot tones. Each terminal node transmits known symbols on two of the four pilot tones, and nulls the signal on the other two pilot tones.

We adopt BPSK and QPSK modulations and the regular Repeat Accumulate (RA) channel code \cite{divsalar1998coding} with code rate  $1/3$. We adapt the virtual channel decoder for PNC developed in \cite{lu2012asynchronous} for our purpose here (see Section III). For each round of virtual channel decoding, we perform 20 BP iterations within the channel decoder. The signal-to-noise ratio (SNR) is defined as ${E_b}/{N_0}$, where ${{E_b}}$ is the energy per source bit.

We evaluate the mean square error (MSE) of ${e^{j{{\widehat\Theta }_{u,m}}}}$ ($E{\left\| {{e^{j{{\widehat\Theta }_{u,m}}}} - {e^{j{\Theta _{u,m}}}}} \right\|^2}$) and the BER of the network-coded messages. We investigate the performance of the proposed EM-BP method for joint phase tracking and channel decoding. We benchmark our method against the traditional method with pilot-based phase tracking and channel decoding. The phase tracking of traditional method employs the LS estimation given by (9).

\subsection{Simulation Results}
We investigate two channel models. The first is the flat Rayleigh fading channel. The second is a frequency-selective Rayleigh fading channel. We model the frequency-selective channel as a tapped delay line channel  ${h_u}\left( t \right) = \sum\nolimits_{l = 0}^{{L_u} - 1} {{\alpha _{u,l}}\delta \left( {t - l{T_s}} \right)}$, where $L_u$  is the number of multipaths, $\alpha _{u,l}$ is the multipath gain. We assume the all multipath gains are independent Rayleigh fading and the power delay profile of the paths satisfies $E\left( {{{\left| {{\alpha _{u,l}}} \right|}^2}} \right) \propto \exp \left( { - cl} \right)$ and $\sum\nolimits_{l = 0}^{{L_u} - 1} {E\left( {{{\left| {{\alpha _{u,l}}} \right|}^2}} \right) = 1}$, where $c$ is the power decay factor. The power decay factor determines the envelope of the frequency-domain channel responses. The larger the power decay factor, the flatter the frequency-domain channel responses (as $c \to \infty$, the channel is reduced to flat fading). In our simulations, we change the value of $c$ to investigate its impact on performance. We set the numbers of multi-paths for nodes A and B to ${L_A} = {L_B} = 4$. The information of the channels is perfectly known in the simulations. For every pair of frames, the normalized CFOs that cause phase drifts on our signals are generated from a uniform distribution over the range $\left[ { - 0.5\delta ,0.5\delta } \right]$, where $\delta $ is the CFO attenuation factor. Since we focus on the residual CFO after precoding, we set a small value of $\delta  = 0.1$.

Fig. 2 presents the MSE results and Fig. 3 presents the BER results under the flat fading channel. From the MSE results for the flat fading channel in Fig. 2, we can clearly see that the EM-BP algorithm gives more accurate channel estimation than the traditional pilot-based phase tracking. The estimation accuracy in EM-BP improves progressively with the number of iterations. As for the BER results of flat fading channel in Fig. 3, we can see that there is a 2 dB gain by EM-BP PNC just after the first EM iteration (K = 1). There is a 3dB gain after EM has converged (K = 7).

\begin{figure}[!t]
\centering
\includegraphics[width=3.5in]{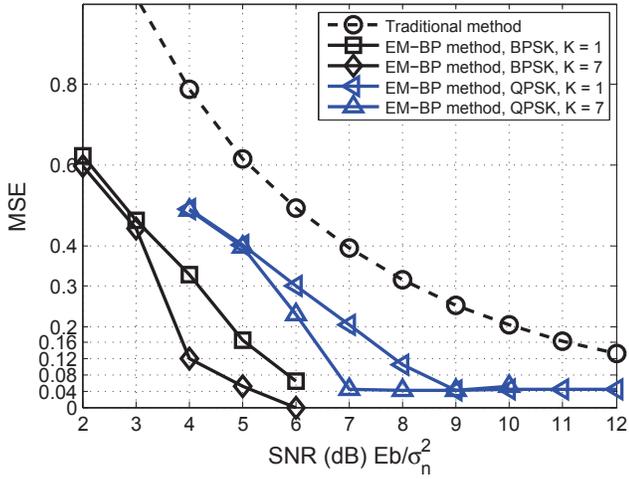}
\caption{The simulated MSE results of EM-BP in frequency-flat channel.}
\end{figure}

\begin{figure}[!t]
\centering
\includegraphics[width=3.5in]{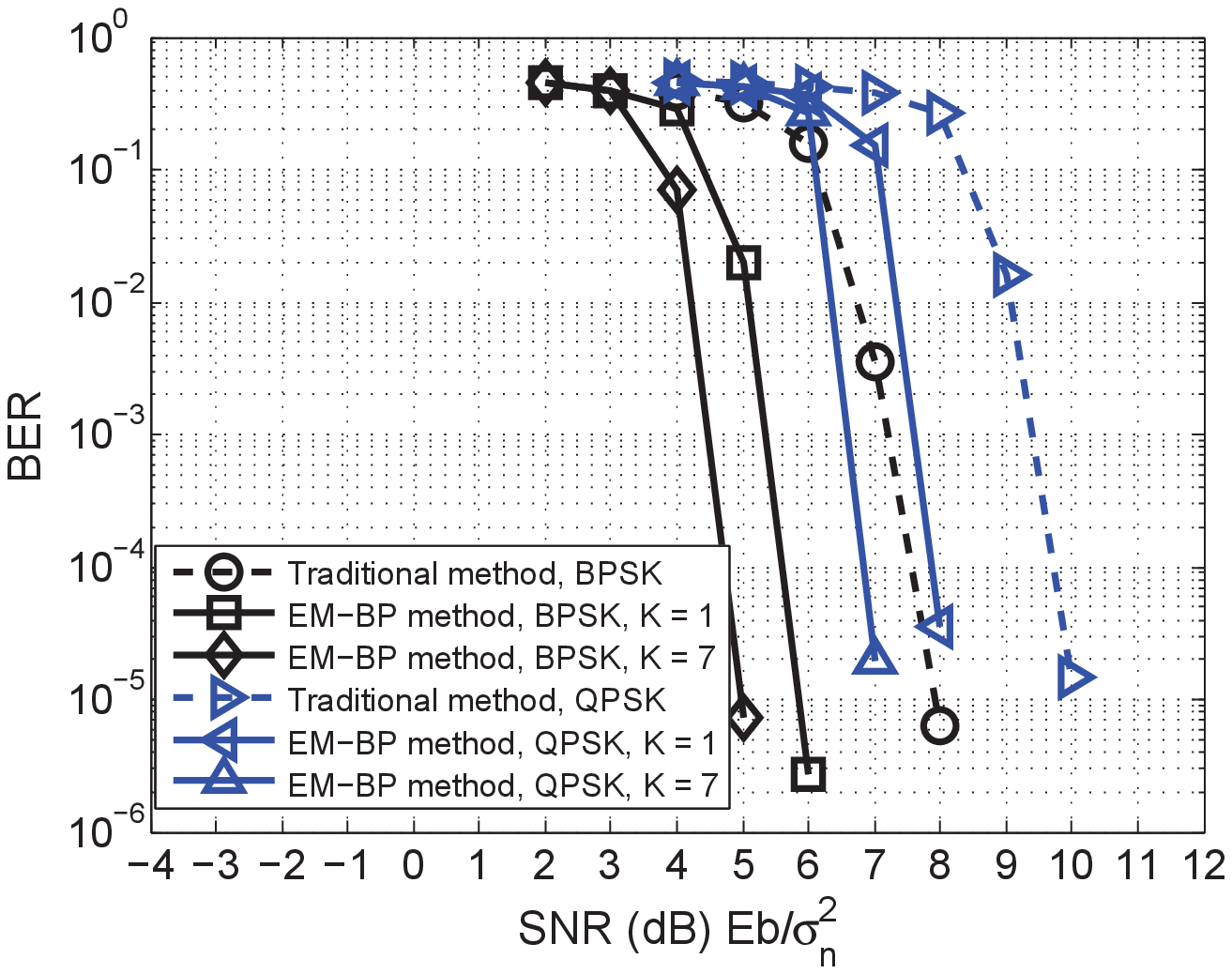}
\caption{The simulated BER results of EM-BP in frequency-flat channel.}
\end{figure}

\begin{figure}[!t]
\centering
\includegraphics[width=3.5in]{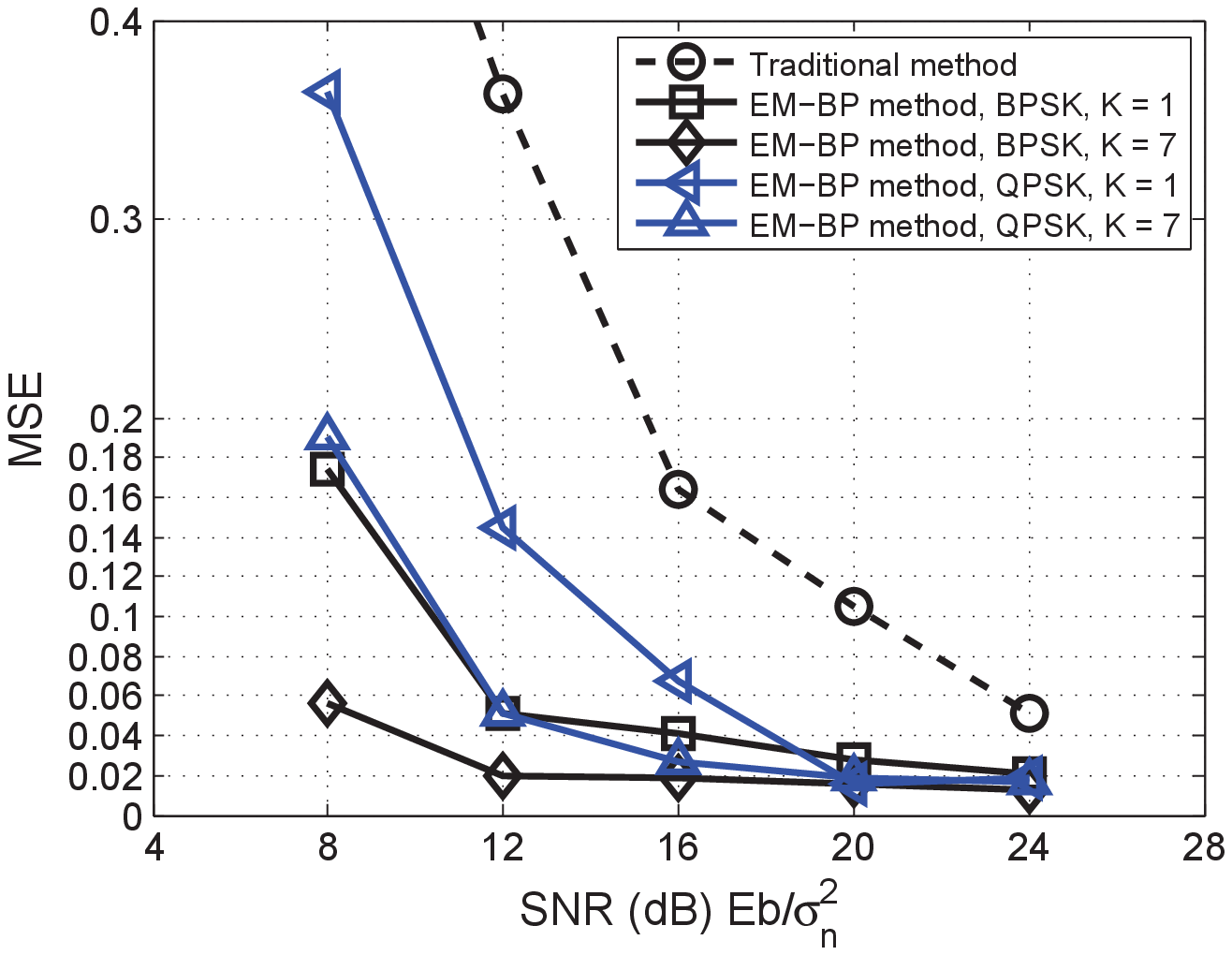}
\caption{The simulated BER results of EM-BP in frequency-flat channel.}
\end{figure}

\begin{figure}[!t]
\centering
\includegraphics[width=3.5in]{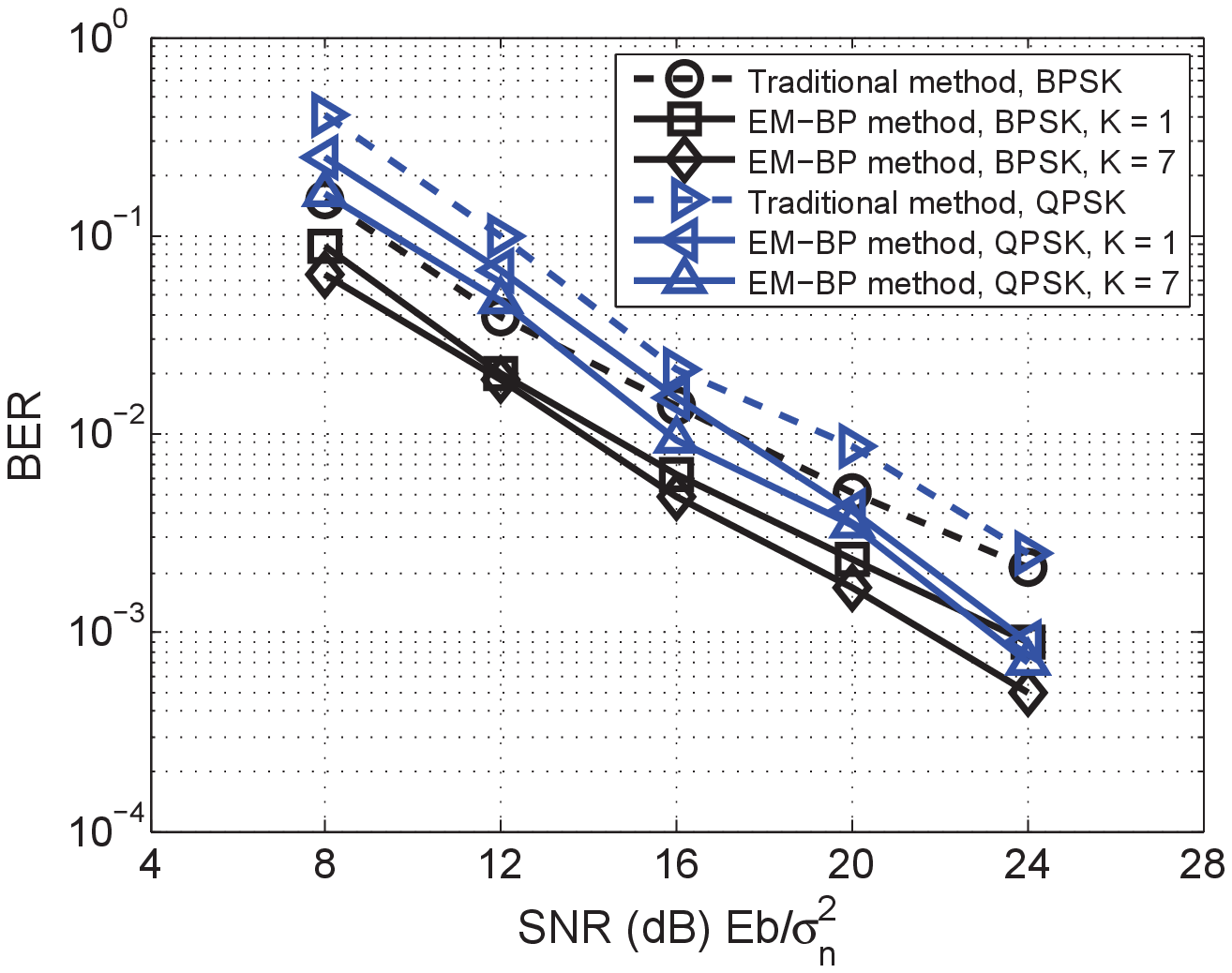}
\caption{The simulated BER results of EM-BP in frequency-selective channel.}
\end{figure}

\begin{figure}[!t]
\centering
\includegraphics[width=3.5in]{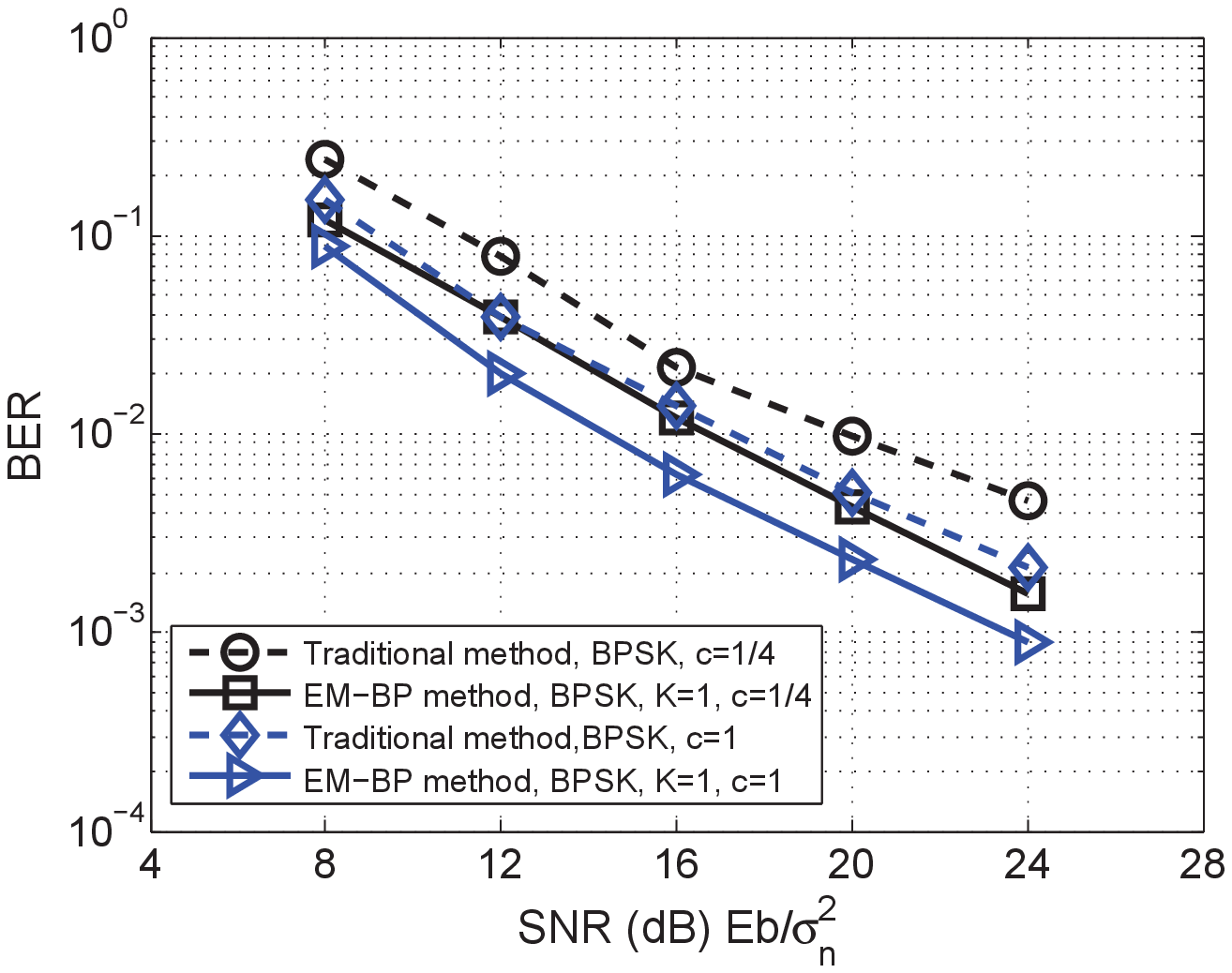}
\caption{The simulated BER results of EM-BP in frequency-selective channel.}
\end{figure}

\begin{figure}[!t]
\centering
\includegraphics[width=3.5in]{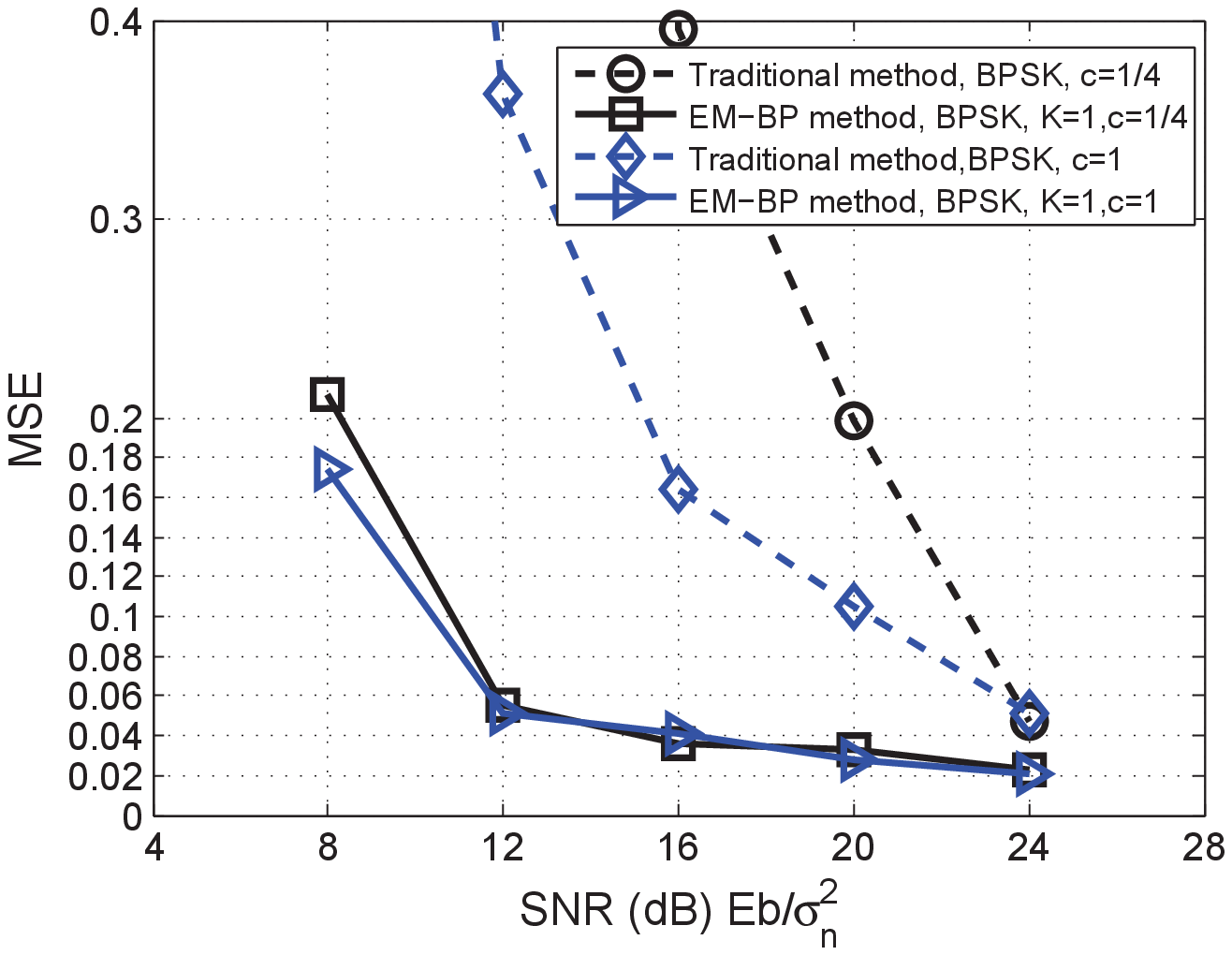}
\caption{The simulated BER results of EM-BP in frequency-selective channel.}
\end{figure}

Fig. 4 and Fig. 5 respectively present the MSE results and the BER results of the frequency-selective channel with the power decay factor $c=1$. We see that the performance trends for the frequency-selective channel are similar to that for the flat fading channel. We also investigate the impact of the power decay factor   on the performance. The BER and MSE results in the frequency selective channel with different power decay factor ($c = 1/4$ and $c=1$) are presented in Fig. 6 and Fig. 7, respectively.  From the results in Fig. 6, we can see that the BERs of the traditional method and the EM-BP method in the channel with larger c are better, since the channel now is flatter. Regardless of $c$, the gain in BER by EM-BP still holds. One interesting observation for the MSE results shown in Fig. 7 is that the MSE of EM-BP is more robust against smaller $c$. With smaller $c$, the channel is more frequency-selective. If the frequency-domain channels on the pilot tones are in deep fading, the pilot based phase-tracking of the traditional method cannot be accurate. Since the EM-BP method employs both data and pilots tones for phase-tacking, it is less affected  by the deep fading of the frequency-domain channels on some tones.

\begin{figure}[!t]
\centering
\includegraphics[width=3.5in]{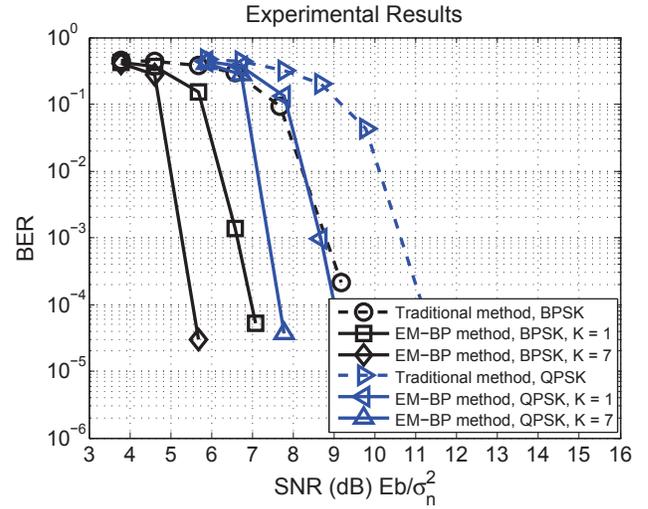}
\caption{The experimental BER results.}
\end{figure}

\subsection{Experiment Results}
Going beyond simulations, we also evaluate the performance of the proposed method experimentally. We use the data collected from a prototype of the OFDM PNC system \cite{lu2013real}. The prototype is built on the USRP N210 hardware \cite{usrp} and the GNU Radio software with the UHD hardware driver \cite{gnu}. To emulate a TWRC system, we deployed three sets of USRP N210 with XCVR2450 boards \cite{gnu} in our lab. The relay node R uses 802.11 channel 1 (2.412GHz) to poll the two end nodes to transmit together at channel 11 (2.462GHz). The system bandwidth is 4MHz. The use of 4 MHz bandwidth rather than the 20MHz full 802.11 bandwidth is due to the limitation of the USRP hardware.

We perform controlled experiments for different SNRs. The receive powers of packets from nodes A and B at the relay are adjusted to be balanced (power imbalance within 1dB). The relay transmits 100 beacons to trigger 100 simultaneous uplink transmissions for each fixed SNR. After the terminal nodes receive the beacon from the relay, they estimate the CFO from the beacon (beacons consist of 2 long training symbols as defined in 802.11 format). Then, they perform CFO precoding on the signals before the uplink transmissions. Finally, the relay receives the simultaneous transmissions from the two nodes and converts it into digital data to be processed by the proposed method. In the experiments, the channels between the relay and the two nodes are estimated using the orthogonal preambles of the packets \cite{lu2012implementation, lu2013real}.

The experimental BER results are shown in Fig. 8. In general, we observe similar performance trends as our simulation results: 2-3 dB gain by the EM-BP method. In particular, the BER performances of our experiments are closer to the flat-fading channel simulations results.  The reason is that the bandwidth used in our experiments (4 MHz) is not large enough for frequency selectivity to come into play.

\section{Conclusion}
We have investigated the use of an EM-BP algorithm to solve the problems of phase tracking and channel decoding in OFDM PNC systems jointly. The main principle of our method is to use the soft information on the data produced by channel decoding to improve the performance of phase tracking, and to use the better phase tracking results to improve channel decoding, in an iterative manner. Our simulation and real experiment results showed that the proposed method can obtain 2 dB gain after the first iteration, and around 3 dB gain after convergence of the iterations.

\ifCLASSOPTIONcaptionsoff
  \newpage
\fi

\bibliographystyle{IEEEtran}
\bibliography{symbollevelcombining}

\end{document}